# Performance Optimization of Surface Electromyography (sEMG) based Biometric Sensing System for both Verification and Identification

Ashirbad Pradhan, Jiayuan He, Member, *IEEE*, and Ning Jiang*, Senior *Member, IEEE*

*Abstract*— Recently, surface electromyography (sEMG) emerged as a novel biometric authentication method. Since EMG system parameters, such as the feature extraction methods and the number of channels, have been known to affect system performances, it is important to investigate these effects on the performance of the sEMG-based biometric system to determine optimal system parameters. In this study, three robust feature extraction methods, Time-domain (TD) feature, Frequency Division Technique (FDT), and Autoregressive (AR) feature, and their combinations were investigated while the number of channels varying from one to eight. For these system parameters, the performance of sixteen static wrist and hand gestures was systematically investigated in two authentication modes: verification and identification. The results from 24 participants showed that the TD features significantly ($p<0.05$) and consistently outperformed FDT and AR features for all channel numbers. The results also showed that the performance of a four-channel setup was not significantly different from those with higher number of channels. The average equal error rate (EER) for a four-channel sEMG verification system was 4% for TD features, 5.3% for FDT features, and 10% for AR features. For an identification system, the average Rank-1 error (R1E) for a four-channel configuration was 3% for TD features, 12.4% for FDT features, and 36.3% for AR features. The electrode position on the flexor carpi ulnaris (FCU) muscle had a critical contribution to the authentication performance. Thus, the combination of the TD feature set and a four-channel sEMG system with one of the electrodes positioned on the FCU are recommended for optimal authentication performance.

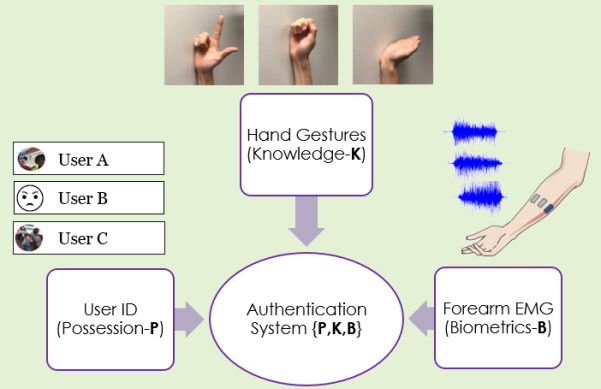

## I. INTRODUCTION

Authentication systems are critical and integral components of modern societies, can be found in personal electronic devices, law enforcement, and airport security [1]. Modern authentication systems usually operate on two modes: knowledge (of a unique code) and biometrics, respectively [2]. The knowledge-based authentication mode usually takes in the form of a unique code, known only to the genuine user. Passwords and PIN (personal identification number) are the most common choices of this authentication mode. On the other hand, the biometrics-based authentication mode utilizes traits and characteristics that are uniquely specific to an individual, such as fingerprints or iris. These two modes can be incorporated separately but also be combined to provide a higher level of security.

This work was supported by the Natural Sciences and Engineering Research Council of Canada (Discovery Grant: 072169).
AP, JYH and NJ are with Engineering Bionics Lab, Department of Systems Design Engineering, Faculty of Engineering, University of Waterloo, Waterloo, Canada.
(*Corresponding author: ning.jiang@uwaterloo.ca)

In the biometric mode, there are two applications: 1) verification and 2) identification. Verification is the application of accepting or rejecting the claim of a user by comparing his/her biometric traits to that of the user's enrollment data stored in the system. Identification is the application of finding the closest match of a presented biometric trait(s) among all the enrollment data stored with the system. In general, verification applications have a higher requirement for the uniqueness of biometric traits than verification applications.

Ideally, a system with both authentication modes (knowledge and biometrics) would be preferred to maximize security, as either mode has inherent drawbacks. One of the advantages of the knowledge-based mode is that the code (password or PIN) can always be regenerated. On the other hand, common biometric traits, such as fingerprint and facial information, cannot be changed once they are compromised, posing a lifetime risk for the genuine user [3]. As such, a system with a dual authentication mode with simultaneously knowledge-based and biometric information is highly desirable.

Recently, various biosignals such as electroencephalogram (EEG), electrocardiogram (ECG), and electromyogram (EMG)

from the brain, heart, and the muscle, respectively have been used as biometric traits [3-6]. These biosignals are suggested to be used as an auxiliary biometric trait because they have two important disadvantages compared to well-developed well-established biometric traits. First of all, the performance accuracy is considerably lower [7]. Secondly, they are less robust against conditions such as illness, physical and mental activities, and the time interval between the training session and the real-time authentication session [8, 9]. As such, research efforts have been undertaken to improve the accuracy and robustness of biometric systems based on biosignals. It is likely that they will find applications in low-risk authentication scenarios, complementary to other well-established biometric modalities. Although multiple studies have investigated EEG and ECG as a biometric trait, there have been limited studies that use surface electromyogram (sEMG) from the forearm and hand muscles while performing hand gestures [6, 10-12]. Due to the characteristic property that different movements result in distinctive EMG patterns, sEMG has been predominantly used for accurate hand gesture recognition [13-15]. However, while training with multiple users, individual differences were observed [16, 17] which motivated the biometric studies based on sEMG. One crucial advantage of sEMG over other biometric traits is its potential for the dual-mode authentication system described above. With sEMG, the knowledge-based information can be implemented by gesture recognition, in which hand/wrist gesture can be used as authentication 'code'. The biometric information is based on the individual characteristics embedded in sEMG. As the concept of using sEMG as an authentication method is relatively new, the existing literature is limited. Initial studies focused on the classification of individuals using different pattern recognition algorithms and only reported the classification accuracy [18-20]. This made the results difficult to establish sEMG as a biometric trait by comparing it to current biometric traits, such as fingerprints, iris, and face. A different sEMG-based study utilized the two applications of biometrics, verification and identification, and reported an equal error rate (EER) of lower than 10% [6]. The study involved 24 participants who performed 16 hand gestures. The frequency division technique (FDT) was used for the feature extraction method [21]. Other studies used a High-density sEMG (HD-EMG) setup where 64 channels were placed in a grid formation on the dorsal side of the hand [10, 11]. The feature extraction involved a combination of time-frequency-spatial domain features as well as HD-EMG decomposition-based features. A recent study utilized only one electrode placed on the flexor digitorum superficialis muscle while the participants performed a phone unlock pattern with contractions of their fingers [12]. The time-domain feature extraction method was employed on the sEMG signals. All these above studies showed a low error rate (<15%) of average authentication performance, indicating sEMG as a promising biometric trait. However, considering many parameters of sEMG processing system, especially the feature extraction methods and the number of electrode channels [22, 23] can change the system performance, it is important to investigate their effects on the authentication performance and to determine an optimal processing configuration of using sEMG based biometric applications.

### A. Feature Extraction Methods:

The three main categories of sEMG feature extraction are: time domain [24], frequency domain [25-27], and time-frequency domain [25, 28, 29]. Additionally, there are non-linear feature extraction methods [30] that adhere to the non-stationary and complex nature of the sEMG signals. However, it has been shown that for a short time contraction, the sEMG signals satisfy the stationary assumptions [31]. Therefore, for a short duration, isometric contractions suitable for biometric applications [10-12], the time domain and frequency domain features are ideal for feature extraction as their mean and variances have minimal variations [32]. These include Hudgin's time-domain (TD) feature set [24] and two frequency domain feature sets: Frequency division technique (FDT) [21, 26] and Autoregressive coefficients (AR) [27, 33]. Time Domain (TD) features are computed using the signal amplitude values. The computed features namely mean absolute value (MAV), slope sign changes (SSC), zero crossing (ZC), and waveform length (WL) are a quantification of sEMG waveform amplitude, frequency, and duration. Due to their relatively lower computational cost, they have been preferred as a feature set in multitudes of pattern recognition-based research [13]. The frequency-domain category of features is based on the principle that the sEMG spectrum varies with limb movements and has shown robustness in real-time myoelectric control of prosthetic functions [34, 35]. The AR feature set includes the coefficients of the nth order autoregressive model of the sEMG signal [27, 33, 36] and has shown to be stable over varying muscle force levels and shifts in electrode location [37]. To the best of our knowledge, there is no existing research on the comparison of feature extraction methods for sEMG based authentication. Therefore, it is crucial to determine the best feature-set that not only has good gesture recognition accuracy but also considers individual differences between users.

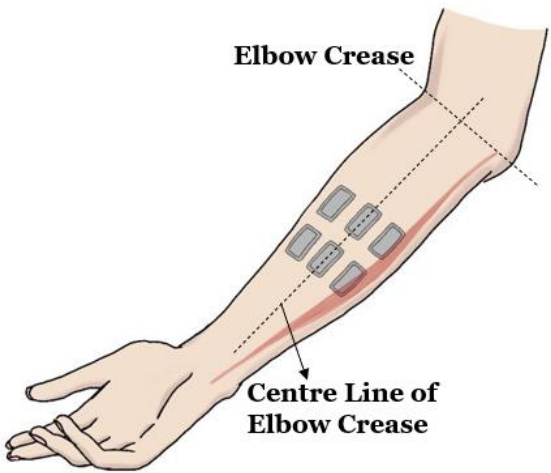

Fig. 1. Positions of the sixteen monopolar electrodes on the forearm (dorsal view). In each ring, eight electrodes (grey rectangles on the forearm) are evenly placed, and one electrode is placed in the centerline of the elbow crease. The distance between the centers of the two rings was 2 cm, and the proximal ring was placed one third of the forearm length between the elbow crease and the upper ring.

### B. Number of Channels

In addition to the feature extraction method, the number of channels is also an important factor affecting the performance of sEMG based gesture recognition. It has been shown that increasing the number of channels, as well as the use of HD grids provides a more accurate gesture recognition [38, 39]. However, space limitation, power consumption, and viability in a practical scenario hinder the use of a large number of sEMG channels. In this regard, previous studies have analyzed the effect of channel reduction on gesture recognition performance [40, 41]. It was shown that four to six channels are sufficient for gesture recognition [41]. It has been shown that a uniform electrode positioning of five electrodes around the forearm, regardless of muscle anatomical location provided a high accuracy of gesture recognition [42]. Therefore, research on channel reduction is important to assess performance in sEMG based authentication systems, thus facilitating the development of easy-to-use wearable sensors. The findings of the study helped determine the best feature set in terms of performance accuracy and computational complexity as well as the optimal number of channels necessary to make accurate authentications using sEMG.

Therefore the purpose of the study was 1) to compare different sEMG feature extraction techniques and their combinations on the authentication performance, 2) to study the effect of channel reduction on the authentication performance, 3) to determine the optimal configuration (feature extraction method and the number of channels) for accurate authentication performance using sEMG based biometrics.

## II. METHODS

### A. Participants

Twenty-four healthy participants (13 males and 11 females, age: 19-30 years) were recruited for the study. The average forearm length (measured from the styloid process to the olecranon) was 25.4 ± 1.46 cm. Participants were informed that they may withdraw from the study at any time during the experiment session. The experiment protocol was in accordance with the Declaration of Helsinki and approved by the Office of Research Ethics of the University of Waterloo (ORE#: 22391). Additionally, data from 20 participants (age: 22-30), publicly available through the NinaPro DataBase DB7, was used in the analysis [43].

### B. Experimental Protocol

The participants were seated comfortably in a height-adjustable chair in an upright position with both of their upper limbs in a resting position and pointing towards the ground. Sixteen monopolar sEMG electrodes (AM-N00S/E, Ambu, Denmark) were placed in the form of two rings, each consisting of eight electrodes equally spaced around the forearm and forming bipolar pairs (Fig. 1). To make the electrode position more uniform across all participants, one of the electrodes in each ring was positioned in the centerline of the elbow crease. The center-to-center distance was 2 cm between the two rings, and the proximal ring was at a distance of one-third the forearm length from the elbow crease. A computer screen placed in front displayed visual instructions to help the participant perform the defined gestures. The sEMG signals were acquired using a commercial amplifier (EMG USB2+, OT Bioelettronica, Italy). The sampling rate was 2048 Hz, and a bandpass setting of the device was set at 10 Hz and 500 Hz. Eight bipolar channels were created from data by taking the differential of each paired electrode from the two rings and used for subsequent processing.

The following sixteen hand and wrist gestures were included in the current study (Fig. 2): lateral prehension (LP), thumb adduction (TA), thumb and little finger opposition (TLFO), thumb and index finger opposition (TIFO), thumb and little finger extension (TLFE), thumb and index finger extension (TIFE), index and middle finger extension (IMFE), little finger extension (LFE), index finger extension (IFE), thumb extension (TE), wrist flexion (WF), wrist extension (WE), forearm supination (FS), forearm pronation (FP), hand open (HO), hand close (HC). Each gesture was repeated seven times and five seconds of recording was performed each repetition. To avoid sEMG signals from transitory movements, the participants were asked to start the contraction a little bit earlier than the beginning of the recording. A five-second rest period was provided between two consecutive contractions to avoid muscle fatigue. The order of the gestures was randomized for each participant. For the public dataset (NinaPro DB7), participants performed 17 isometric and isotonic, hand and finger gestures. Six repetitions of each gesture were performed and 5 seconds of data were recorded using a Delsys Trigno IM Wireless EMG system [43].

### C. Signal Processing and Feature Extraction Methods

The bipolar surface EMG signals were first windowed with a window length of 200 ms and an overlap of 150 ms, *i.e.* 50 ms overlap between two consecutive windows. Each window was processed subsequently using three commonly used EMG feature extraction techniques: 1) Hudgin's Time Domain (TD) feature extraction [24], 2) Frequency Division Technique [21], and 3) the autoregression (AR) technique [27]. For the TD feature set, time-domain features (mean absolute value, zero crossing, slope sign changes, and waveform length) were extracted from filtered data. The mathematical form of these features for the signal $x$ is provided below

$$RMS = \sqrt{\tfrac{1}{N}\sum_{i=1}^{N} x_i^2} \tag{1}$$

$$SSC = \sum_{i=2}^{N-1} f(x_i - x_{i-1})(x_i - x_{i+1})$$
$$f(x) = \begin{cases} 1, & \text{if } x \geq Th \\ 0, & \text{otherwise} \end{cases} \tag{2}$$

$$WL = \sum_{i=1}^{N-1} |x_{i-1} - x_i| \tag{3}$$

$$ZC = \sum_{i=1}^{N} sgn(x_i \times x_{i+1}) \cap (x_i - x_{i+1})x \geq Th$$
$$sgn(x) = \begin{cases} 1, & \text{if } x \geq Th \\ 0, & \text{otherwise} \end{cases} \tag{4}$$

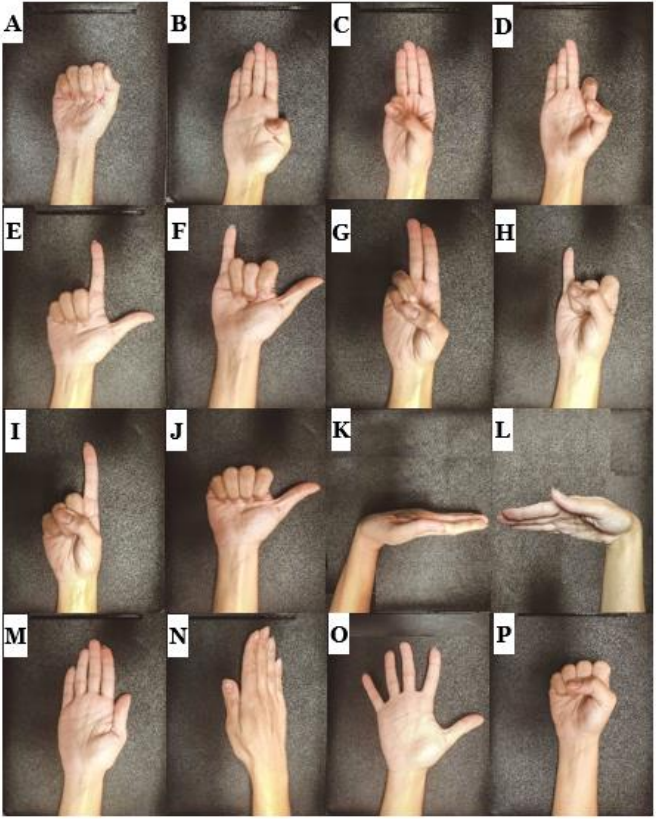

Fig. 2. Sixteen gesture classes investigated in the study: (A) lateral prehension (LP), (B) thumb adduction (TA), (C) thumb and little finger opposition (TLFO), (D) thumb and index finger opposition (TIFO), (E) thumb and little finger extension (TLFE), (F) thumb and index finger extension (TIFE), (G) index and middle finger extension (IMFE), (H) little finger extension (LFE), (I) index finger extension (IFE), (J) thumb extension (TE), (K) wrist flexion (WF), (L) wrist extension (WE), (M) forearm supination (FS), (N) forearm pronation (FP), (O) hand open (HO), (P) hand close (HC).

For FDT, the signals from each channel were divided into specific frequency sub-bands [21] as given by:

$$FDT_i = F\left[\sum_{j=1}^{N_i} |X(f_{i,j})|\right], \quad i = 1,2...L \quad (5)$$

where $F(\cdot)$ is a nonlinear transformation, normally logarithm, to ensure the smoothness of the values.

AR model is a prediction model defined as:

For the AR feature set, the 6th order autoregression coefficients ($a_p$) were used for it was widely adopted in previous studies [27]. Therefore, the three different feature vectors (TD, FDT, and AR) were analyzed separately for authentication performance. Additionally, the combination of the time domain and frequency domain features was investigated using the following feature sets: TD+FDT and TD+AR. For a given feature vector sample $p$ (the input), its similarity score, namely $S_{i,j}$, with the $i$th gesture and the $j$th user, was defined as the Mahalanobis distance between the sample and the class centroid is

$$S_{i,j}(p) = \sqrt{(p - \mu_{i,j})^\top \Sigma_{i,j}^{-1}(p - \mu_{i,j})} \quad (7)$$

where $\mu_{i,j}$ is the centroid of the gesture $i$th class and the $j$th user and $\sum_{i,j}$ is the covariance matrix for the specific gesture and user class Both the parameters are calculated from the system training data and the sample $p$ is from the system testing data. The similarity score was used to compare the true user with the imposter. The leave-one-out (LOO) cross-validation scheme was used, where six trials were used for training and one trial for testing. The performance evaluation was

$$x_i = \sum_{p=1}^{P} a_p x_{i-p} + w_i \quad (6)$$

performed separately for the two biometrics applications: verification and identification.

### D. Channel selection and reduced channel set

In the current study, eight bipolar channels resulting from the two rings of electrodes placed on the forearm. This hardware configuration might not be feasible for more practical applications in which a smaller number of electrodes as possible should be used. A channel selection algorithm based on the sequential forward selection (SFS) method [citation] was implemented to examine system performance when only a subset of the available sEMG channels is used. The procedure of the SFS is described below. Two channel-sets were defined: the applied channel set **A** (initially empty, i.e. $\mathbf{A} = \phi, m = 0$) and the remaining electrode set **R** (initially full, i.e. $\mathbf{R} = 1, 2, ..., N, n = N$). The number $m$ is the size of **A**, $n$ is the size of the **R**, and $N$ is the total number of available channels (8 in the current study).

In each iteration, a union of A and one of the channels from **R**, i.e. $r_k$, was used for feature extraction and subsequent authentication system evaluation. This step is repeated for all channels in **R**. Then, the channel that produces the minimum EER in the case of the verification system and minimum RkE (k=1,5) in the case of the identification system was chosen, respectively. These errors are discussed in the following sections. Thus, for the $j$th iteration:

$$\begin{aligned} r_j &= \arg\max_{r_i} \{EER_{A\cup\{r_i\}} : \\ EER_{A\cup\{r_i\}} &= f_{EER}(A \cup \{r_i\}) \\ for\ r &= 1, 2...R(n)\} \end{aligned} \quad (8)$$

$$\begin{aligned} r_j &= \arg\max_{r_i} \{RkE_{A\cup\{r_i\}} : \\ RkE_{A\cup\{r_i\}} &= f_{RkE}(A \cup \{r_i\}) \\ k &= 1, 5\ \&\ for\ r = 1, 2...R(n)\} \end{aligned} \quad (9)$$

The applied set and remaining set are updated as

$$\begin{aligned} A &= A \cup \{r_j\},\ R = R \setminus \{r_j\}, \\ m &= m+1,\ n = n-1,\ for\ j = 1, 2...N \end{aligned} \quad (10)$$

The iterations are repeated till **R** is empty and **A** contains all $N$ channels. The authentication system evaluation parameters (EER and RkE, k=1,5) in each iteration stage are reported for analysis. Also, the range namely $r$, which is the difference between the maximum and minimum error for every iteration is reported to analyze the distribution of error. Therefore, $EER_r$ for the verification systems R1E$_r$ and R5E$_r$ for the identification system were reported, respectively. The error ranges are a measure of electrode location significance (for a particular iteration of channel selection, the higher the error range, the more significant is the specific channel selected).

### E. Performance Evaluation of the Verification System

The performance analysis of the verification system was performed using the detection error tradeoff curve (DEC), where false rejection rate (FRR) was plotted against false acceptance rate (FAR) for various Mahalanobis distance thresholds (for accepting or rejecting the claim). FRR represents the likelihood of a legitimate request from the true user is rejected. FAR represents the likelihood of illegitimate access requests (from imposter) is granted. Two standard metrics were calculated from the DEC: the area under the curve (AUC) and the equal error rate (EER). AUC refers to the area under the DEC [2] and EER refers to the point on the DEC curve where FAR is equal to FRR [2]. The lower the AUC and EER value the better the authentication performance.

'A code' in the current framework is a gesture performed by a user (either the authorized user or an imposter). Therefore, to facilitate subsequent presentation and discussion, the two words 'code' and 'gesture' are used equivalently. This nomenclature is also followed for the analysis of the identification system. In the current study, the code length was fixed at one gesture. All the codes were evaluated individually. Assuming a user uses a specific gesture for authentication, which is called the authentication gesture, three possible scenarios would arise in the verification system: 1) Normal Test, in which the authentication gesture was known only to the authorized user, and the imposter attempts to guess the authentication gesture and gain access, *i.e.* the code was a secrete and not compromised; 2) Leaked Test, in which the authentication gesture was known to an imposter who tries to use it to gain access, *i.e.* the authentication code was compromised; and 3) Self Test, in which the authorized user forgets the authentication gesture and tries to gain access by performing other known gestures to the system. In all three scenarios, the genuine data is the data from the authentication gesture of the authorized user. For the Normal Test scenario, the imposter data was the data from all the other gestures and all the other users. For the Leaked Test scenario, the imposter data was the data from the authentication gesture of all other users. Lastly, for the Self Test scenario, the imposter data was the data from the other gestures of the true user.

### F. Performance Evaluation of the Identification System

In the literature, the authentication performance of the identification system was quantified by the *rank-k* error [6]. The *rank-k* error represents the likelihood of the authorized user is not among the top *k* users returned by the identification system. In this study, the smaller the similarity score is, the higher it would be ranked by the identification system. The *rank-k* error is a decreasing function with respect to the value of *k* as a property of the cumulative match characteristic curve, which plots *rank-k* error against *k*. In this study, *rank-1* error (R1E), which is a crucial index when only one code is used, was reported for this study. Additionally, the *rank-5* error (R5E) was reported for comparison with previous studies [6]. The cumulative match characteristics (CMC) curve was plotted for the optimal configuration of sEMG system.

### G. Statistical Analysis

The purpose of this study was to determine the best feature extraction method and the optimal number of channels for both the biometric authentication systems (verification and the identification) based on sEMG. The non-parametric Kruskal Wallis test was employed to determine if the two factors, feature extraction methods (five levels, *i.e.* FDT, TD, and AR, TD+FDT, TD+AR) and the number of channels (eight levels, *i.e.* from 1 to 8), have significant effects on the performance metrics, EER, AUC, R1E, and R5E. In case of significance in either of the two factors, the level of the other factor was fixed, and the Wilcoxon rank-sum test was performed on the other factor.

For the optimal configuration, performance metrics were reported for the best feature extraction method and the optimal number of channels (for both verification and identification). The performance of optimal parameters were then compared with the NinaPro DB7. All statistical tests were performed using RStudio 1.0.136 (RStudio, Boston, MA).

## III. RESULTS

### A. Performance Evaluation of Verification System

Fig. 3 shows the distribution of the median EER and the median AUC across all participants, respectively for the three feature extraction methods (FDT, TD, and AR), their combinations (TD+FDT and TD+AR), and a different number of channels (1-8). Median EER for the FDT method and for eight channels was 0.76% (Q1=0.1%, Q3=2.93%) for the Normal Test, 2.8% (Q1=0.3%, Q3=5.45%) for the Leaked Test, and 4.86% (Q1=2.1%, Q3=9.89%) for the Self Test (where Q1 is the 25$^{th}$ percentile and Q3 is the 75$^{th}$ percentile). While using a TD feature extraction method for the eight channels, the median EER reduced to 0.27% (Q1=0.002%, Q3=2.47%) for the Normal Test and 0.48% (Q1=0.001%, Q3=3.51%) for the Leaked Test. The Self Test using the TD feature set resulted in a median error of 6.68% (Q1=2.76%, Q3=14.04%). The combined feature sets along with TD feature set are illustrated in Fig. 4. For the Leaked Test scenario, it was observed that the median EER for TD+FDT and TD+AR was 0.16% (Q1=10$^{-4}$%, Q3=0.25%) and 0.18% (Q1=10$^{-4}$%, Q3=0.25%). The AUC followed a very similar pattern to the EER for scenarios (reported in the next paragraph). When the number of channels was higher than four, the verification system's EER reached a plateau. These values are similar to other verification systems reported in the literature [6]. For the three feature extraction methods (FDT, TD, and AR) the initial channel (n=1) EER$_r$ was higher than 0.05 for all the scenarios. The EER$_r$ using the TD and FDT method decreased with an increasing number of channels and it reached a plateau (EER$_r$ <0.01) when the number of channels was higher than two. A similar plateau (EER$_r$ <0.01) was reached for the AR method with the number of channels higher than three (n>3). The combined feature sets

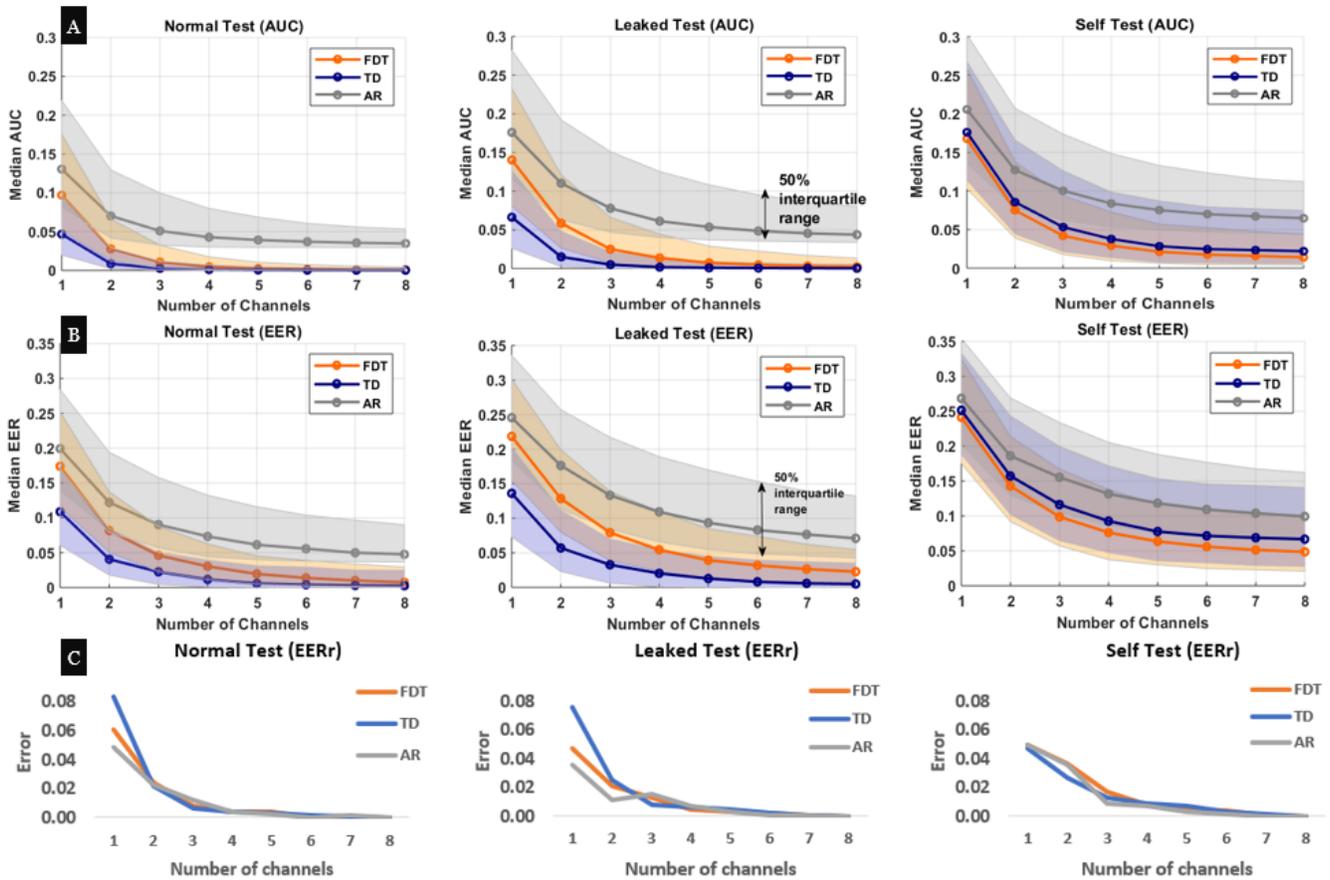

**Fig. 3. (A and B)** The median AUC and the median EER for the three feature extraction methods (FDT, TD and AR) and number of channels (1-8) for the Normal Test (left column), Leaked Test (middle column) and Self Test (right column) scenario of the verification system. Each line indicates the respective median values, and shaded area indicates the corresponding middle 50 percentile (25% to 75%). **(C)** The range of EER values ($EER_r$) for the feature extraction methods (FDT, TD and AR) and number of channels (1-8) for the Normal Test, Leaked Test, and Self Test scenario of the verification system. The range is calculated as a part of the sequential forward selection algorithm at the end of every iteration.

(TD+FDT and TD+AR) had similar $EER_r$ to the TD feature sets for all number of channels (n=1-8).

**Feature Extraction Methods:** In the Normal Test scenario, the median EER was significantly lower for TD than both FDT and AR methods ($p<0.05$, Fig. 3(A)) regardless of channel numbers. The AR method had the highest AUC than the other two methods ($p<0.05$). Similarly, the median AUC for the TD method was significantly lower than FDT, which was in turn significantly lower than AR ($p<0.05$). In the Leaked Test, the EER and AUC both had similar outcomes as the Normal Test: TD outperforms FDT, which in turn outperformed AR ($p<0.05$, Fig. 3(B)). For Self Test, the EER and AUC were not significantly different between the TD and FDT methods ($p>0.05$, Fig. 3(C)) regardless of the channel numbers. However, other than the single-channel case, the EER and AUC of AR were significantly higher than the other two methods ($p<0.05$). For all three scenarios, the $EER_r$ of the TD method was higher than FDT and AR for lower channel lengths (1-2) (Fig. 3(C)). The $EER_r$ was very similar for all the methods for channel lengths (3-8). For a combination of TD with spectral features it was observed that TD+FDT and TD+AR had similar values in all the three scenarios and for different channel lengths ($p>0.05$). It was observed that for the Normal Test and Leaked Test, TD+FDT and TD+AR had a significantly lower ($p<0.05$) EER than the TD feature set. However, the median EER values were similar for the three methods: for a 4-channel configuration, the EER for TD method was 0.2% (Q1=$10^{-3}$%, Q3=0.54%), the EER for TD+FDT was 0.1% (Q1=$10^{-3}$%, Q3=0.42%) and EER for TD+AR was 0.1% (Q1=$10^{-3}$%, Q3=0.42%). For the Self Test scenario, there was no significant difference ($p>0.05$) between the three feature sets for all number of channels.

**Number of Channels:** As expected, there was a general decreasing trend in both EER and AUC values with increasing channel numbers (Fig. 3) in all the scenarios. For the Leaked Test, the decrease in AUC was significant from one to four channels ($p<0.05$), for all feature extraction methods. The decrease was not significant when more channels were included (5-8, $p>0.05$). The EER had similar results, with a significant decrease in the lower number of channels (1-4, $p<0.05$) for FDT and AR methods. However, for the TD method, the decrease in EER was only significant from one channel to three channels ($p<0.05$, Fig. 3(A)). Similar outcomes were observed in the Normal Test scenario for all three methods. When the channel numbers were smaller, there was a significant decrease with increasing channel number: the AUC and EER decreased significantly for smaller channel numbers (1-4, $p<0.05$) for TD and FDT methods and from one to three ($p<0.05$) for the AR method. For the Self Test scenario, the AUC and EER significantly decreased when the channel number increased from one to two ($p<0.05$) for all the three feature extraction methods, and the decrease was not significant for the channel numbers (2 to 8, $p>0.05$). For all the feature extraction methods the EER$_r$ decreased from channel length 1 to 8, consistently across all analysis scenarios (Fig. 3(C)). The decrease in range was minimal (EER$_r$) when the number of channels was between three and eight.

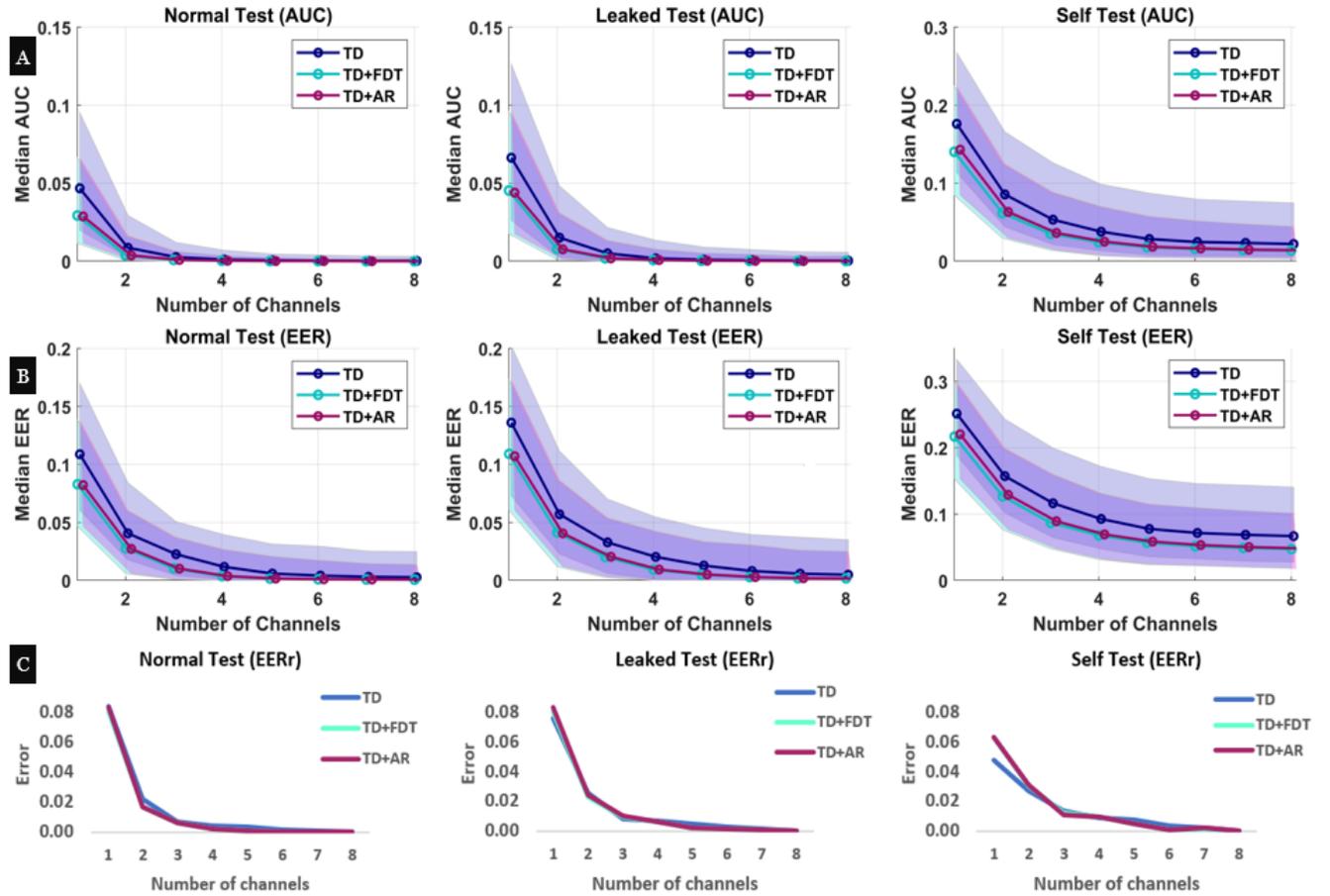

**Fig. 4.** (A and B) The median AUC and the median EER for the combined feature sets (TD, TD+FDT and TD+AR) and number of channels (1-8) for the Normal Test (left column), Leaked Test (middle column) and Self Test (right column) scenario of the verification system. Each line indicates the respective median values, and shaded area indicates the corresponding middle 50 percentile (25% to 75%). (C)The range of EER values (EER$_r$) for the feature sets (TD, TD+FDT and TD+AR) and number of channels (1-8) for the Normal Test, Leaked Test, and Self Test scenario of the verification system. The range is calculated as a part of the sequential forward selection algorithm at the end of every iteration.

### B. Performance Evaluation of Identification System

R1E and R5E were used to quantify the authentication performance with different feature extraction methods and channel numbers. Fig. 5(A) shows the distribution of the R1E and R5E for the different methods and the number of channels. Median R1E for the FDT method and eight channels was 2% (Q1=0%, Q3=12%) and Median R5E was 0% (Q1=0%, Q3=0.2%) for the identification system, which was in agreement with a previous study having a similar experimental setup. The TD feature set had a comparatively lower median R1E of 0.2% (Q1 = 0%, Q3= 4.6%) and a similar R5E of 0% (Q1=0%, Q3=0.1%). The AR feature set had a median R1E of 18.8% (Q1=7.4%, Q3=57.9%). For the combined feature sets, the median R1E for TD+FDT was 0.06% (Q1 = 0%, Q3= 2.9%) and median R1E for TD+FDT was 0.2% (Q1 = 0%, Q3= 4.6%). It was evident that the R1E reached a plateau (median R1E≤3%) when the number of channels used was four and higher. Fig. 5(b) shows the R1E$_r$ and R5E$_r$ for the feature extraction methods and the number of channels. Both the R1E$_r$ and R5E$_r$ decreased with an increasing number of channels. For the TD feature extraction method, the R1E$_r$ for one channel was 0.17 and the R1E$_r$ for two channels was 0.09. Thereafter, the R15$_r$ was lower than 0.02 and reached a plateau for more than three channels. The FDT had an R1E$_r$ of 0.076 (n=1), 0.072 (n=2) and 0.05 (n=3) and thereafter it plateaued with R1E$_r$<0.015. The AR had an R1E$_r$ of 0.057 (n=1), 0.045 (n=2) and 0.028 (n=3) and reached a plateau for more than three channels. The combined feature sets (TD+FDT and TD+AR) had R1E$_r$ similar to the TD feature sets for number of channels higher than two (n>2). The R5E$_r$ showed a similar distribution for the feature extraction methods and the number of channels.

**Feature Extraction Methods:** The R1E was significantly lower for the TD method ($p<0.05$) than FDT and AR for all the number of channels (n=1-8). Also, the FDT was significantly lower ($p<0.05$) than AR for all the channels. For the R5E analysis, the difference between the feature extraction methods

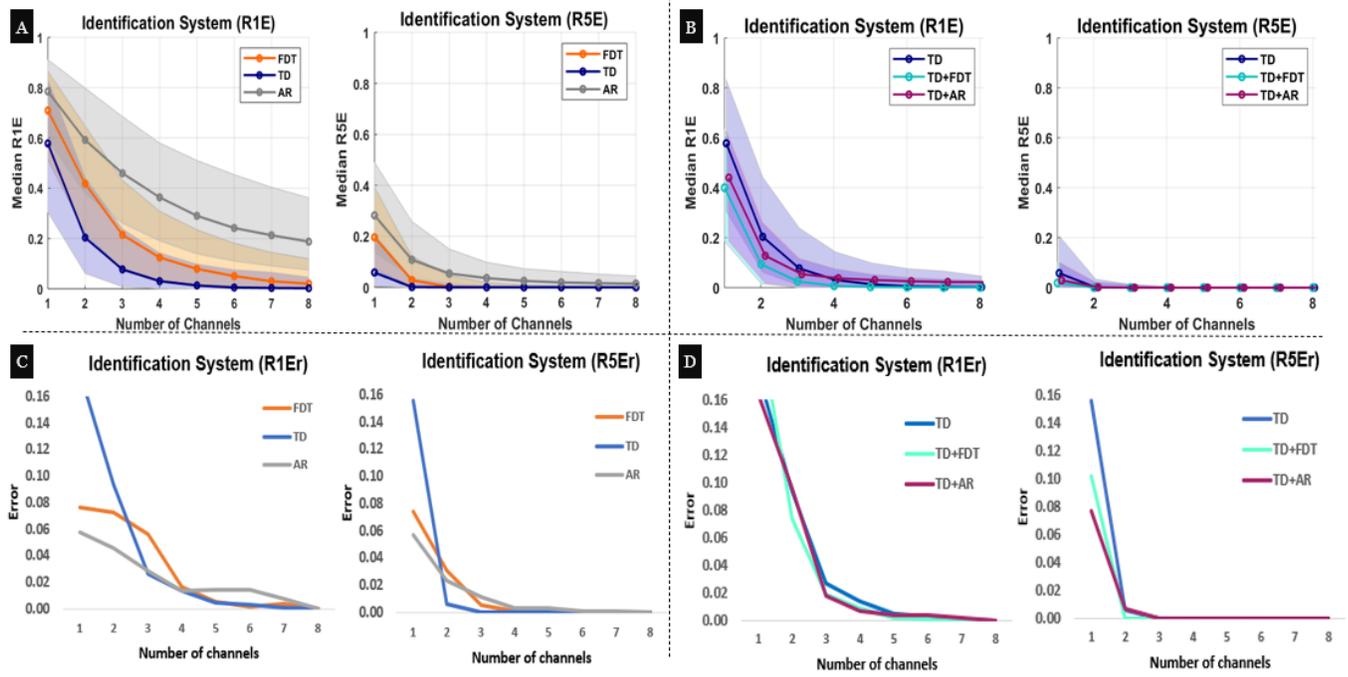

**Fig. 5.** (A and B) The median R1E and the median R5E for the processing methods (FDT, TD, AR), their combinations (TD, TD+FDT, TD+AR) and number of channels (1-8) for the identification system. Each line indicates the respective median values, and shaded area indicates the corresponding middle 50 percentile (25% to 75%). (C and D) The range of the R1E (R1E$_r$) and R5E (R5E$_r$) for the feature extraction methods (FDT, TD and AR) and the combined feature sets (TD, TD+FDT,TD+AR) for different number of channels (1-8) for the identification system. The range is calculated as a part of the sequential forward selection algorithm at the end of every iteration

(TD, FDT, and AR) was seen for the lower number of channels (1-3). The TD method had a significantly lower R5E for 1-2 number of channels. However, for the inclusion of more channels (3-8), there was no difference between the R5E for TD and FDT, but the AR had a significantly higher R5E than TD and FDT ($p<0.05$). For a combination of TD with spectral features it was observed that the R1E for TD+FDT was significantly lower ($p<0.05$) than the TD+AR and TD for different channel lengths. For a 4-channel configuration, the EER for TD+FDT was 0.7% (Q1=$10^{-4}$%, Q3=7.58%), which was significantly ($p<0.05$) lower than the R1E for TD method (Median=3%, Q1=$10^{-3}$%, Q3=14.3%), and R1E for TD+AR (Median=3.8% (Q1=$10^{-3}$%, Q3=11.2 %). The R1E$_r$ was higher for the TD method for the number of channels one and two (1-2) and was the lowest from channel length (3-8, Fig. 5(C)). The R5E$_r$ was the highest for the TD method for the first channel number and was the lowest from 2-8. The R1E$_r$ was higher for the FDT method than the AR method for a lower number of channels (1-3) and thereafter the R1E$_r$ dropped below 0.02 for both the methods.

**Number of Channels:** The R1E for the TD method decreased significantly for the number of channels one to four (1-4, $p<0.05$). There was no significant decrease in further addition. For the FDT and AR method, the R1E significantly decreased for the number of channels one to three (1-3, $p<0.05$). The R5E significantly decreased for all three methods ($p<0.05$) for channels 1-2. From 3-8 the error was almost constant with the TD and FDT having a median R1E of zero. For all the methods, the R1E$_r$ decreased from the number of channels 1-8, however, the decrease was minimal from channels 4-8. For all the methods, the R5E$_r$ also decreased from the number of channels 1-8, however, the R5E$_r$ plateaued from channels 3-8.

For both the verification and identification system (the optimal parameters (TD Feature selection and 4 electrode channels, as discussed in section IV.(C)) were used to test the performance of NinaPro DB7 (shown in Table I). The results show that the EER for Normal Test was 6.4% (Q1=3.4%, Q3=11.2%), the EER for Leaked Test was 6.8% (Q1=3.2%, Q3=11.9%) and the EER for Self Test was 23.5% (Q1=18.2%, Q3=29.7%). The R1E for the identification system was 10.9% (Q1=4.5%, Q3=21.9%).

## IV. DISCUSSION

This paper systematically investigated the effects of feature extraction methods and the channel reduction using EMG to achieve optimal authentication performance in the two application systems: user verification and user identification. The main advantage of using EMG over other biosignals is the high accuracy of dual-mode authentication with both knowledge-based and biometrics-based authentication information. The ability to use sEMG for recognizing accurate gestures allows the user to choose unique gestures as their own authentication code. In the verification system, the Normal Test scenario fully utilizes this dual-mode authentication property of sEMG. The Normal Test scenario represents the authentication system with both the layers of protection (knowledge-based and biometrics-based). In the Leaked Test scenario, only

TABLE I
TD FEATURE EXTRACTION AND
FOUR-CHANNEL EMG CONFIGURATION

| Verification (EER) | Current Study | | | NinaPro Database | | |
|---|---|---|---|---|---|---|
| | Med | Q1 | Q3 | Med | Q1 | Q3 |
| Normal Test | 0.012 | 0.001 | 0.040 | 0.064 | 0.034 | 0.112 |
| Leaked Test | 0.020 | 0.001 | 0.055 | 0.068 | 0.032 | 0.119 |
| Self Test | 0.093 | 0.048 | 0.172 | 0.235 | 0.182 | 0.297 |

| Identification (R1E) | Current Study | | | NinaPro Database | | |
|---|---|---|---|---|---|---|
| | Med | Q1 | Q3 | Med | Q1 | Q3 |
| | 0.030 | 0.001 | 0.143 | 0.109 | 0.045 | 0.219 |

TD: Time domain feature set, EER: Equal error rate,
R1E: Rank-1 Error, Q1: 1st quartile (25%) Q3: 3rd quartile (75%)

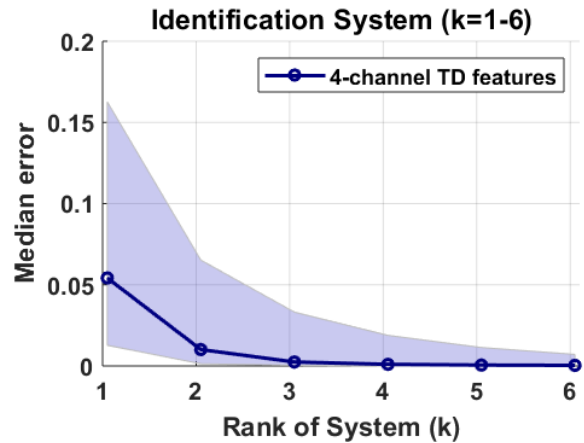

Fig. 6. CMC curve for the identification system using 4 channels sEMG and TD Feature extraction.

biometric-based protection is available as the code is compromised. And the Self Test scenario further illustrated the unique authentication power of the sEMG-based system, in which the authentication code is forgotten but the legitimate user can be still verified by performing other gestures (codes). With a typical combination of four-channel and FTD feature set, the median EER of the Leaded Test (5.4%) was higher than that of the Normal Test (3%), indicating a 2.4% additional protection provided by the user-defined authentication code (Fig. 3 (A and B)). The nature of the verification system makes it challenging to get a low error (specifically R1E). The genuine user must be always predicted as rank-1 or else it is considered an error. The median R1E for a four-channel system was 12.4% (FDT feature set), 3% (TD feature set), and 36.4% (AR feature set). Clearly, the TD feature set had the best authentication performance which is discussed in the next section.

### A. Feature extraction methods

The TD feature extraction method resulted in better authentication performance than FDT and AR for two scenarios of the verification system (the Leaked test and Normal Test, shown in Fig. 3) and the identification system (Fig. 5). For the Self Test, both the TD and FDT had similar performance across all channel numbers and outperformed the AR method, which had the poorest performance in all number of channels. This was in agreement with the previous studies that compared AR with the TD and FDT methods [44]. For the identification system, the TD feature set had a significantly lower median R1E (3%) than that of the FDT (12.4%). Therefore, the TD feature set has an overall superiority in performance over FDT for both verification and identification systems. This outcome suggests that the TD feature set is likely more sensitive to individual-differences in sEMG, crucial for biometric applications. Previous studies developing a generalized multiuser model found that common spatial-spectral analysis (CSSA) and spectral moments, both similar to FDT, have higher accuracy than the TD features [17, 45]. These studies suggest that these features are less user-dependent, which means they capture fewer individual characteristics than TD features, and might explain the finding in this study that the TD feature performing better than FDT and AR feature sets.

While comparing TD with the combined feature sets, the average EER of TD (4%) was higher than TD+FDT (2.67%) and TD+AR (2.7%) for the verification system, and the average R1E for TD (3%) was higher than TD+FDT (0.7%) and lower than TD+AR(3.8%) for the identification system. The difference in performance for the TD feature set, although significant ($p<0.05$), was considered acceptable while comparing to other biometric traits [7]. Due to the superior performance of the TD feature set and the relatively lesser computational complexity than the combined features, it was considered as the optimal feature extraction technique for biometric applications.

The median EER for the TD feature set and an eight-channel setup was 0.27% (Normal Test), 0.48% (Leaked Test), lower than those reported by a different study (Leaked Test EER = 14.96%) which used 64 channels and a combined feature set of (sample entropy, spectral entropy, median frequency, waveform length, and root mean square) [10]. This corroborates our findings and suggests that TD feature set would have superior performance than other feature extraction methods even with a disadvantage of the number of channels.

### B. Number of channels and channel reduction

As expected, the median EER reduced while increasing the number of channels for all scenarios of the verification system and for the identification system. This is in accordance with the previous gesture recognition studies that investigated channel reduction [22, 40, 41]. It is evident from Fig. 3 (A,B) and Fig. 5 (A,B), a four-channel setup was considered to provide a balance between overall authentication performance and complexity (*i.e.* channel numbers). The error values for the TD feature set are comparable to those reported in previous sEMG biometric-related studies [6, 10, 12]. A previous study used 64 channel HDEMG [10]. Our results indicated that there might not be a need for high-density EMG as four-channel setup provided similar authentication performance. While a one-channel (on the FDS muscle) configuration was used in [12], the half total error rate (HTER) of less than 90% was not sufficient for any practical authentication system.

The error range parameters $EER_r$, $R1E_r$, and $R5E_r$ were used in the study to determine if the channel location has a significant impact. For all the scenarios it was observed that the first electrode has a higher error range than the remaining channels as shown in Fig. 3 (C) and Fig. 5 (C and D). In the sequential forward selection, the lowest error channel was selected in every iteration. Therefore, for the one channel setup (first iteration) the higher error range suggests a significant impact on the first channel location. As observed in all the scenarios, the electrode placed on the Flexor Carpi Ulnaris (FCU) consistently contributed the most towards the authentication performance. This was in agreement with the study that determined the best forearm muscle to control a soft robotic glove [46]. From channels 2-8, the impact was shown to be lower, thus indicating the remaining electrodes could be placed irrespective of specific muscle locations, a desirable feature in practical applications.

### C. Optimal configuration.

The EER and R1E for the four-channel setup and TD feature extraction are listed in Table I. From the above findings, it can be seen that the combination of a four-channel sEMG setup, of which one is attached on FCU, and the TD feature set, the average EER for the Normal Test (1.2%), Leaked Test (2%) and Self Test (9.3%) can satisfy the requirement of a verification system (shown in Table I). The identification system R1E was 3% for this setup. These values are comparable to another study that reported an error of 0.5%, however, they used 64 channels and a combination of different features [11]. Another study used an eight-channel forearm bracelet and a combination of frequency domain and time domain features and obtained a FAR of 0.2% and FRR of 2.9% [47]. A different study using an eight-channel setup and FDT feature set reported an R1E error of 9% [6]. In the present study, a very low R1E of 0.2% can be achieved while using an 8-channel configuration and TD feature extraction. Fig. 6 presents the cumulative match characteristics (CMC) curve for such an identification system. It was observed that with an increase in rank (k) of the system, the recognition error (RkE) reduced, thus improving the performance. The analysis was repeated for the NinaPro DB7 database consisting of 20 participants and performing 17 hand and finger gestures [43]. It was found that for a similar set of parameters (4 channels and TD feature set) the performance was slightly lower than the present study for the Leaked Test (EER=6.4%) and Normal Test (EER=6.8%, reported in Table II). The lower performance observed with the Self Test (EER=23.5%) and Identification System (R1E=10.9%) might be due to the partly isotonic nature of the hand gestures, suggesting further investigation of feature sets specific for dynamic contractions. Therefore, the optimal configuration for an sEMG authentication system consists of four sEMG channels and a TD feature set. If there is a higher performance requirement, using 8-channel setup will provide enhanced performance. The findings of the study helped determine the best feature set in terms of performance accuracy and computational complexity as well the optimal number of channels necessary to make accurate authentications using sEMG.

The performance reported in Table II was comparable to the other commonly used biometric traits (fingerprint, iris, and facial recognition) reviewed in a previous study [7]. The sEMG based biometrics has two similar functionalities as ECG and EEG: hidden nature and liveness detection. The sEMG possesses unique knowledge-based security which allows users to customize their access codes as different gestures. This feature, although not available in ECG can be achieved using EEG by using different mental states for codes. As only accuracy >70% has been achieved for classifying two mental states [48], this is almost impossible for real-world applications. The Leaked Test verification which is a measure of the biometric-based security achieved a median EER of 2%. This was comparable to the reported EER of ECG and EEG biometrics which was 0.1% to 5% [3], and 1% to 20% [4, 5]. For the Identification system, the median R1E was 3% which was in range with the R1E reported for ECG and EEG, 0%-20% [3] and 1% to 20% [4, 5], respectively. This suggests that the four-channel sEMG with TD feature extraction can achieve comparable if not better biometric performance than other biosignals.

### D. Limitation and Future work

While discussed earlier in a previous study, a combination of gestures may significantly improve authentication performance. It is important to note that the results presented here only used one gesture, *i.e.* a single code. Including more than more code will certainly further improve the system performance, which is the topic of a subsequent study. Also, the present study involved data from 24 participants. A significantly larger participant pool will be used for future studies. Another limitation of the present study was the effect of time was not investigated, as all the sessions were performed on the same day, partly due to the COVID constraints currently in place at the University of Waterloo. A multi-session study will be performed to validate the robustness of an sEMG based authentication system.

## V. CONCLUSION

The TD features had improved authentication performance than FDT features and AR features for the Leaked test and Normal test of User Verification System and for all number of channels (n=1:8). The TD features had better performance than the FDT feature set in the User Identification System and for all number of channels (n=1:8). For all the feature sets, four channels had a plateau in performance for a higher number of channels for both the verification and identification system. Therefore, the TD feature set and the four-channel sEMG configuration is found to be ideal for optimal authentication performance using sEMG systems. The range of error was high for the first channel and reduced for the inclusion of further channels. This indicates that one location on the FCU is important and the remaining could be placed irrespective of muscle location. This finding motivates the design of an easy-to-wear arm-band with four equally spaced electrodes and the placement is recommended in a way that one of the electrodes is right on top of the FCU for best authentication performance. Therefore, an accurate sEMG based authentication system for biometric applications can be designed using a TD feature extraction method and four equi-spaced electrodes with one of them positioned directly on the FCU muscle.